\documentclass[twocolumn,showpacs,preprintnumbers,amsmath,amssymb,prl]{revtex4}
\usepackage{epsfig}

\begin{document}

\title{Universality of Brunnian ($N$-body Borromean) four and 
five-body systems}

\author{M.T. Yamashita}
\affiliation{Instituto de F\'\i sica Te\'orica, UNESP - 
Univ Estadual Paulista, C.P. 70532-2, CEP 01156-970, 
S\~ao Paulo, SP, Brazil}
\author{D.V. Fedorov}
\author{A.S. Jensen}
\affiliation{Department of Physics and Astronomy, Aarhus University, 
DK-8000 Aarhus C, Denmark}

\date{\today}

\begin{abstract}
We compute binding energies and root mean square radii for weakly
bound systems of $N=4$ and $5$ identical bosons.  Ground and first 
excited states of an $N$-body system appear below the threshold 
for binding the system with $N-1$ particles.  Their root mean square 
radii approach constants in the limit of weak binding. Their probability 
distributions are on average located in non-classical regions of space 
which result in universal structures.  Radii decrease with increasing 
particle number. The ground states for more than five particles are 
probably non-universal whereas excited states may be universal.
\end{abstract}


\maketitle

\paragraph*{Introduction.}

The Efimov effect was predicted in \cite{efi70} as the appearance of a
series of intimately related excited three-body states when at least
two scattering lengths are infinitely large.  These states can appear
at all length scales and their properties are independent of the details
of the potentials. This effect has in recent years been studied
intensively and extended to a wider group of physical phenomena,
beginning to be known as Efimov physics.  In general we define Efimov
physics as quantum physics where Universality and Scale Invariance
apply.  Universality means independence of the shape of the
interparticle potential.  Scale Invariance means independence of the
length scale of the system.  These conditions are rather restrictive
but a number of systems are known to exist within this window
\cite{rii92,fed94,nie98,jen04,yam03,yam03a}.  The great advantage is
that one theory is sufficient to explain properties without any
detailed knowledge of the interactions \cite{amo97}.  Furthermore,
properties in different subfields of physics are described as
manifestations of the same underlying theory.

Our physical definition of scale invariance originates from the halo
physics first realized and discussed in nuclei \cite{han87}, but
quickly observed as applicable also to small molecules like the helium
trimers \cite{jen04}. This original definition of scale invariance, 
that the concept applies to any length scale is obviously continuous
as exemplified by nuclei, atoms and molecules.  Often the notion of
scale invariance is used in a different mathematical sense where the
spatial extension of the structures in one given system repeats itself
in discrete steps like the factor 22.7 for identical particles
\cite{efi70}.  This is a result of the independence of potential
details and here precisely defining our meaning with the notion of
universality.  As the concepts can be defined in different ways we
will use throughout the paper this original physical meaning of scale
invariance. Together, these two concepts constitute our meaning of
Efimov Physics which to the best of our knowledge has been left
undefined in all previous publications.

The range of validity for such a global theory is only well described
for two and three particles \cite{rii92,jen04,bra06}. For $N=4$ two
states were found in the zero-range, inherently universal, effective
field model \cite{ham07}. These states also appeared as universal in
finite-range models in connection with each Efimov state \cite{ste09}.
This is in contrast to \cite{yam06} where the disentanglement of the 
scales used to regularize the three and four-body zero-range Faddeev-Yakubovsky 
equations gives rise to a dependence of the four-body ground state on 
interaction details. Then a four-body scale is needed in analogy to the 
three-body scale appearing independently on top of the two-body 
properties. This apparent discrepancy between Refs. \cite{ham07,ste09} 
and \cite{yam06} is not yet resolved.

Recently, three experiments evidenced two four-body bound states
connected to an Efimov trimer \cite{Fer09,Zac09,Pol09} in accordance
with the theoretical predictions of Ref. \cite{ste09}.  In two of
these experiments were also observed deviations from universality
\cite{Zac09,Pol09}. Surprisingly, the greatest deviation were observed
for large scattering lengths ($\rightarrow\pm\infty$) - exactly at the
region where universality should apply \cite{Pol09}.  This requires a
theoretical explanation where something should be added in the universal 
model.

Very little is known for five particles with complete solutions
containing all correlations as dictated by the interaction.  With
specific assumptions about only $s$-waves and essentially no
correlations it was concluded in \cite{rii00,jen04} that ground state
halos cannot exist for $N>3$. These assumptions are rather extreme and
could be wrong or only partly correct.  However, if halos exist they
have universal structures as the $N=4$ states obtained in
\cite{ham07,ste09}. These results can only be reconciled by 
wrong assumptions in the halo discussion or by impermissible
comparison between halo ground states and excited states.

It was concluded in \cite{ama71} that Efimov states do not exist for
$N>3$ and furthermore for three particles exist only for dimensions
between $2.3$ and $3.8$ \cite{nie01}. However, by restricting to
two-body correlations within the $N$-body system, a series of (highly)
excited $N$-body states were found with the characteristic Efimov
scaling of energies and radii \cite{sor02,tho07}.  Whether they
maintain their identity and the universal character, when more
correlations are allowed in the solutions, remains to be seen.

Two limits to the universality are apparent.  The first appears for large
binding energy where the resulting small radii locate the system
within the range of the potentials and sensitivity to details must
appear.  The less strict second limit is for excitation energies above
the threshold for binding subsystems with fewer particles.  Structures
with such energies are necessarily continuum states which may, or more
often may not, be classified as universal states depending on their
structures and the final states reached after the decay.

Even for four particles where universality is found
\cite{ham07,ste09}, a number of questions are still unanswered.  For
five and more particles the information becomes very scarce.  A novel
study claiming universality for ground states of a Van der Waals
potential has appeared for particle number less than 40 \cite{han06}.
The critical mass is found as a substitute for the critical strength,
but the computed radii at threshold cannot be reliably extracted.

The purpose of the present paper is to explore the window for Efimov
physics. We shall investigate the boundaries for universality
preferentially leading to general conclusions applicable to systems of
$N$-particles.  We first discuss qualitative features and basic
properties, then extract numerical results for $4$ and $5$ particles
very close to thresholds of binding, and relate to classically allowed
regions.  We only investigate Brunnian ($N$-body Borromean) systems
\cite{bru92} where no subsystem is bound.

\paragraph*{Qualitative considerations.}

For two particles the infinite scattering length corresponds to a
bound state at zero energy.  Variation of $1/a$ around zero produces
either a bound state of spatial extension $a$ or a continuum state
corresponding to spatial configurations correlated over the radius
$a$.

For three particles the Efimov effect appears, i.e. for the same
interaction, $a= \pm \infty$ ($1/a=0$), infinitely many three-body
bound states emerge with progressively smaller binding and
correspondingly larger radii \cite{efi70}.  The ratios of two and
three-body threshold strengths for several potentials were derived in
\cite{han06,ric94,goy95}.  These thresholds for binding one
state can be characterized by a value of $1/a$ \cite{bra06,ste09}.
Infinitely many bound three-body states appear one by one as $1/a$ is
changed from the three-body threshold for binding to the threshold for
two-body binding $1/a=0$. Moving opposite by decreasing the attraction
these states one by one cease to be bound.  They move into the
continuum and continue as resonances \cite{bri04}. For asymmetric
systems with a bound two-body subsystem the three-body bound state
passes through the particle-dimer threshold becoming a virtual state
\cite{yam02}. This behavior holds even for particles with different
masses \cite{yam08}.
  
All three-body $s$-wave states from a certain energy and up are
universal.  However, this is not an {\it a priori} obvious conclusion
but nevertheless true because two effects work together, i.e., for
$1/a=0$ the system is large for the excited Efimov states and for
finite $1/a$ the binding is weak and the radius diverges with binding
\cite{jen04}. Both Efimov states and weakly bound states are much
larger than the range of the interaction.  The continuous connection
of these bound states and resonances is therefore also in the
universal region.

The recent results for four particles were that each three-body state
has two four-body states attached with larger binding energy
\cite{ham07,ste09}.  These four-body states are both described as
having universal features unambiguously related to the corresponding
three-body states for interactions of both positive and negative
scattering lengths.  Detailed information of structure, correlations,
and posssible limits to universality are not available.

The one-to-one correspondence between the two four-body states and one
three-body state can perhaps be extended such that two weakly bound
$N$-body states appear below the ground state of the $(N-1)$-body
state. This seems to be rather systematic for $N$-body Efimov states
obtained with only two-body correlations \cite{sor02,tho07}. If these
$N$-body Efimov states remain after extension of the Hilbert space to
allow all correlations, we can expect these sequences to be continued
to the thresholds for binding by decreasing the attraction.
However, ground and lowest excited states may be outside the universal
region but the sequences may still exist.  In any case the scaling
properties are different for the $N$-body Efimov states in
\cite{sor02,tho07} and the universal four-body states in \cite{ham07,ste09}.

The basic reason for the difficulties in finding detailed and general
answers is related to the fact that the thresholds for binding are
moving monotonously towards less attraction with $N$
\cite{ric94,goy95}. For $N=2$ weak binding and large scattering length
is synonymous. Already for $N=3$ this connection is broken but the
weak binding still causes the size to diverge \cite{jen04}.  The
indications are that for $N>3$ the size remains finite even in the
limit of zero binding.

\paragraph*{Basic properties.}

We consider a system of $N$ identical bosons each of mass $m$.  They
are confined by a harmonic trap of frequency $\omega_t$ corresponding
to a length parameter $b^2_t=\hbar/(m\omega_t)$. The particles
interact pairwise through a potential $V$ of short range $b\ll
b_t$. We shall use the gaussian shape $V= V_0 \exp(-r^2/b^2)$.  The
chosen values of $V_0$, $b$, and $m$ lead to a two-body scattering
length $a$ and an effective range $R_e$. The solution to the
Schr\"{o}dinger equation is approximately found by the stochastic
variational method \cite{suz98}. The results are energies and root
mean square radii.

For two-body systems we know that the $n'$th radial moment only
diverge at threshold of binding when the angular momentum $l \leq
(n+1)/2$, see \cite{jen04}. The equality sign implies a logarithmic
divergence with binding $B_2$ in contrast to the normal power law
$B_2^{l-n/2-1/2}$. For the mean square radius this implies divergence
for $l \leq 3/2$.  For an $N$-body system with all contributions
entirely from $s$-waves we can generalize these rigorous results from
two-body systems \cite{jen04}. The number of degrees of freedom is
$f=3(N-1)$ and the generalized centrifugal barrier is obtained with an
effective angular momentum $l^*=(f-3)/2$.  Divergent root mean square
radius is then expected when $l^*\leq 3/2$ or equivalently when $f\leq
6$ or $N\leq 3$. If this result holds, four-body systems should
have finite root mean square radii even at the threshold of binding.

The size of the system is measured by the square root of the mean
square radius, $\langle$~$r^2/b^2$~$\rangle$, which is expressed in
units of the ``natural'' size of the systems, i.e. the range of the
binding potential.  The dimensionless unit of the binding energy $B_N$
of the system is $\bar B=mb^2 |B_N| /\hbar^2$.  Both Universality and
Scale Invariance is therefore detected by inspection of these
quantities as functions of parameters and shape of the potentials. In
regions where the curves are proportional, we conclude that the
properties are universal and scale invariant.  Results for different
potential shapes can be expressed in terms of a standard potential by
scaling the range. Then the individual curves would fall on top of
each other in the universal regime.

\paragraph*{Clasical allowed region.}

Universal properties can intuitively only appear when the structures
are outside the potentials because otherwise any small modification
would have an effect on the wavefunction.  Consequently the property
would be dependent on these details in conflict with the assumption of
universality.  For two-body systems the relative wavefunction is
therefore universal only if the largest probability is found outside
the potential. This means that this classically forbidden region is
occupied.  The system is extremely quantum mechanical and very far
from obeying the laws of classical physics.

To investigate the relation between universality and the classical
forbidden regions for $N$ particles we need to compare features of
universality with occupation of classical forbidden regions.  For
two-body systems this is straightforward since the coordinate of the
wavefunction and the potential is the same.  The probability of
finding the system where the energy is smaller than the potential
energy is then easy to compute as a simple spatial integral over
absolute square of the wavefunction.

For more than two particles the problem is well defined but the
classically forbidden regions (total energy is smaller than the
potential energy) themselves are difficult to locate.  We attempt a
crude estimate which at best can only be valid on average.  The energy
is computed by adding kinetic and potential energy, i.e.
\begin{eqnarray} \label{e40}
 - B_N = N \langle t_1\rangle + \frac{1}{2} N(N-1) \langle V_{12}
 \rangle \;,
\end{eqnarray}
where we choose an arbitrary particle $1$ to get the kinetic part and
a set of particles $1$ and $2$ to get the potential energy. The
classical region is defined by having positive kinetic energy. For a
two-body gaussian potential we then obtain an estimate of an average,
$r_{cl}$, for the classical radius from
\begin{eqnarray} \label{e50}
\langle V_{12} \rangle > - \frac{2B_N}{N(N-1)} =  V_0 \exp(-r_{cl}^2/b^2)\;. 
\end{eqnarray}

If the distance between two particles is larger than $r_{cl}$ we
should be in the universal region. This value can then be compared to
the size obtained from the average distance between two particles,
$\langle r_{12}^2 \rangle $, computed in the $N$-body system from the
mean square radius \cite{sor02}, i.e.
\begin{eqnarray} \label{e80}
 \langle \frac{r^2}{b^2} \rangle = \frac{N-1}{2N}
 \langle \frac{r_{12}^2}{b^2} \rangle   \;.
\end{eqnarray}
Thus in the classical forbidden region $r_{cl}^2/b^2$ from Eq.(\ref{e50})
should be smaller than $\langle r_{12}^2/b^2 \rangle$ from
Eq.(\ref{e80}).

\paragraph*{The four-body system.}

\begin{figure}[!h]
\vspace*{0cm}
\epsfig{file=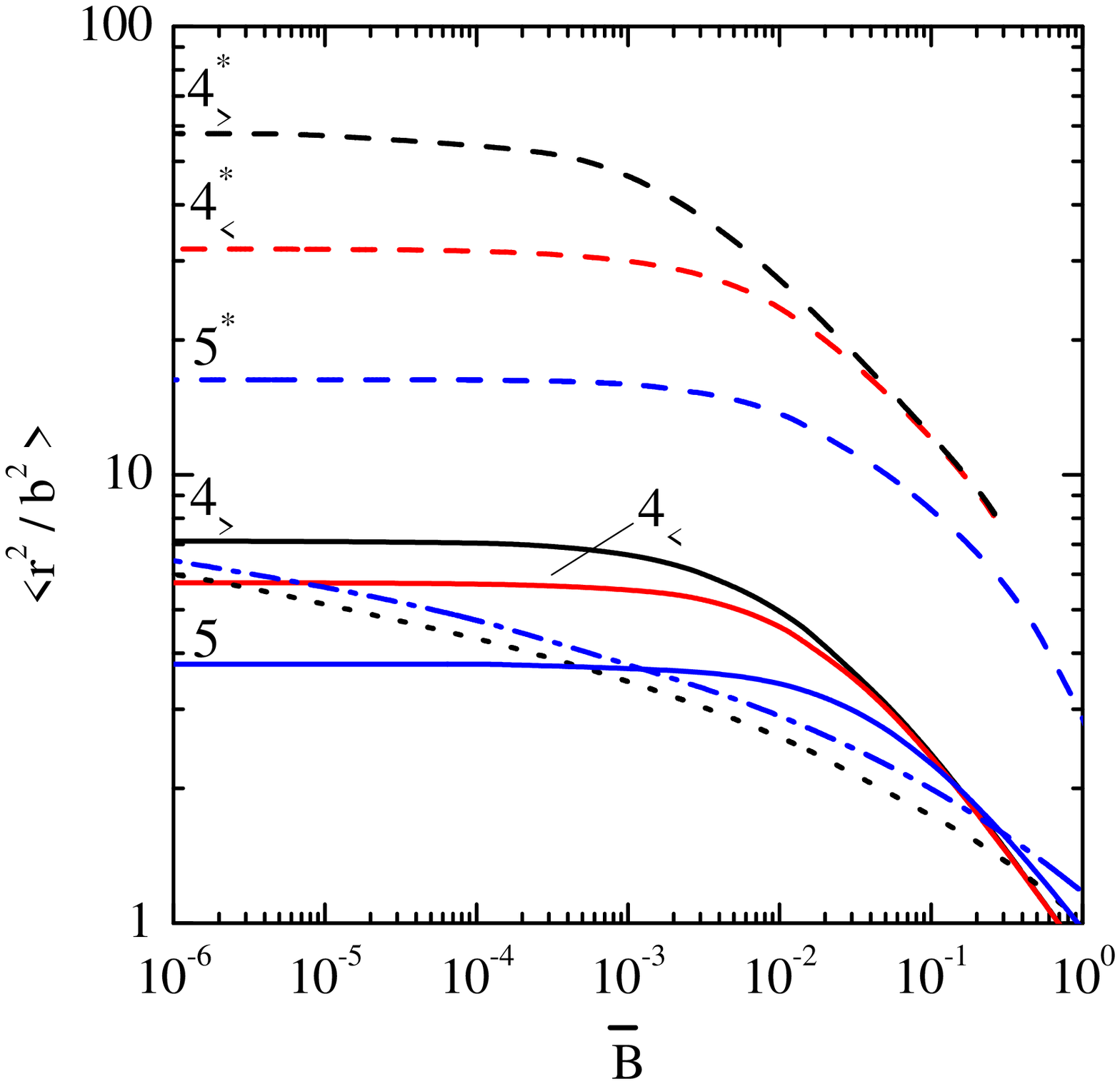,width=8.cm, angle=0}
\caption{(Color online) The mean square radius as function of the four 
and five-body binding energies, all in dimensionless units.  The trap 
sizes for four particles are $b_t= 230.942 a_0$ (red lines with $4_<$), 
$2630.956 a_0$ (black lines with $4_>$), and $b_t= 372.073 a_0$ 
(blue lines with 5) for five particles. Here $a_0$ is the Bohr 
radius.  We show ground (solid) and excited states (long-dashed with 
particle numbers tagged with an *), and ``classical'' 
two-body radius (dotted (4) and dot-dashed (5)) 
translated by Eq.(\ref{e80}).}
\label{fig1}
\end{figure}

We show size versus binding energy for $N=4$ in Fig.\ref{fig1}. The
variation arises by change of the strength, $V_0$, of the attractive
gaussian. The system is for numerical convenience confined by an
external one-body field. However, we are only interested in structures
independent of that field, i.e. intrinsic properties of the four-body
system. We therefore increase the trap size until the states are
converged and located at distances much smaller than the confining
walls.  We now know that this happens for four particles in contrast
to the three-body system where the size diverges when the binding
energy approaches zero.

In Fig.\ref{fig1} we show results for two trap sizes deviating by an
order of magnitude and larger than the interaction range $b=11.65a_0$
($a_0$ is the Bohr radius) by a factor of 20 and 200, respectively.
For large binding in the lower right corner the results for the ground
state is independent of trap size. When the probability extends by
about a factor of $2$ further out than $b$ the effect of the small
trap can be seen. The tail of the distribution then extends out to $20
b$ even though the mean square is 10 times smaller.

In the limit of very small binding energy the radius approaches a
constant independent of the binding. The trap size has to be increased
to $200 b$ before the trap has no influence which implies that the
probability distribution is entirely within that distance when the
threshold for zero binding is reached.  The converged size is about
$2.7 b$ for the ground state. Somewhat surprisingly also the first
excited state, which also is below the energy of the three-body state,
has converged to a value, $7.6 b$, independent of the trap size.  A
shape different from a gaussian would again lead to constants related
through specific properties of the potentials, but the ratio would
remain unchanged. This is precisely as found in two dimensions for
three particles \cite{nie97}.  Both states are at the threshold on
average very much smaller than both traps. Nevertheless the smallest
trap would still influence the tail of the distribution.

In Fig.\ref{fig1} we also show the estimated classical average
distance between pairs of particles within the $N$-body system. This
curve is above the ground state radius for large binding. Here the
probability is mostly found in the classical region within the
potential, i.e. in the non-universal region.  Another potential shape
would then move these curves.  The classical and root mean square
radius cross each other when the size is slightly larger than the
range $b$. This limit for universality is similar to the halo
condition for universality established in \cite{rii00,jen04}.  At
smaller binding energy the classical radius becomes less than the size
of the system and the probability is on average located outside the
potential in the non-classical, universal region.

For the extremely small binding energies close to the threshold our
estimate of the classical radius diverges logarithmically with binding
energy. Thus at some point it has to exceed the size of the system
which we concluded converge to a finite value for zero binding. This
is simply due to the character of the gaussian potential which
approaches zero for large radii.  Zero energy must then be matched by
an infinite radius.  However, this gaussian tail is too small to
obstruct the convergence of the probability distribution to a finite
size.  This cannot destroy universality because the tail has no
influence on the wavefunction in this region far outside the range of
the potential.  For universality only the binding energy is decisive
as one can see explicitly for the two-body system.  For $N$-body
systems the same result follows from the asymptotic large distance
behavior of the wavefunction expressed in hyperspherical coordinates
\cite{mer76}. Thus the classical average radius argument fails for
these extreme energies when the probability has settled outside the
range of the short-range potential.

\paragraph*{Five-body system.}

In Fig.\ref{fig1} we also show results for $N=5$ where convergence is
reached for the trap size of $b_t = 372 a_0$.  We found two pentamers 
with energies -0.0281$\frac{\hbar^2}{mb^2}$ and -0.0113$\frac{\hbar^2}
{mb^2}$ below the four-body threshold (-0.0103$\frac{\hbar^2}{mb^2}$) 
for infinite scattering length. The sizes for both ground
and excited states increase again with decreasing binding energy $B_5$
and approach finite values when $B_5=0$.  These limiting radii of
about $1.94b$ and $4.0b$ are substantially smaller than corresponding
values for four particles. Still the largest probability is found
outside the potential providing the binding.  This strongly indicates
that also these structures are in the universal region.  Again their
ratio is anticipated to be essentially independent of potential
shape. This conclusion is supported by the comparison in
Fig.\ref{fig1} to the classical radius which always is smaller than
the radius of the excited state and comparable to the radius of the
ground state.  As argued for four particles the largest binding for
the ground state corresponds to non-universal structure. When the
binding energy is about $0.3$ in the dimensionless units on the figure
the universal structure appears.  This happens at about the same
energy as for four particles. In both cases the probability is pushed
outside the potential and universality is expected for smaller binding
energies.

\paragraph*{Conclusions.}

We have investigated the behavior of Brunnian systems near threshold 
for binding.  Ground and first excited state for four and five
identical bosons appear below the threshold for binding three and four
particles, respectively. Their radii are for small binding energies
larger than the range of the potential holding them together.  The
largest part of the probability is found in non-classical regions
resulting in universal structures. For six and more particles the
ground states would be located inside the potential and thus of
non-universal structures.  Excited states are larger and may still be
universal but already for seven or eight particles also the first
excited state is expected to be non-universal.  The numerical results
are obtained for a two-body gaussian potential but the features
originating from wavefunctions in non-classical regions of space are
expected to be independent of the potential shape.

\paragraph*{Acknowledgement.}

We are greatful for helpful discussions with M. Th{\o}gersen. MTY
thanks FAPESP and CNPq for partial support.

\end{document}